\documentclass[10pt,letterpaper]{article}
\usepackage{opex3}
\usepackage{amssymb,amsmath}
\usepackage{cite}
\begin{document}

%%%%%%%%%%%%%%%%%% title page information %%%%%%%%%%%%%%%%%%
\title{Exact polarizability and plasmon resonances of partly buried nanowires}

\author{Jesper Jung$^{*}$ and Thomas G. Pedersen}

\address{Department of Physics and Nanotechnology, Aalborg
University, Skjernvej 4A, DK-9220 Aalborg {\O}st, Denmark and
Interdisciplinary Nanoscience Center (iNANO), Denmark}

\email{$^{*}$jung@nano.aau.dk} %% email address is required



\begin{abstract}
The electrostatic polarizability for both vertical and horizontal
polarization of two conjoined half-cylinders partly buried in a
substrate is derived in an analytical closed-form expression. Using
the derived analytical polarizabilities we analyze the localized
surface plasmon resonances of three important metal nanowire
configurations: (1) a half-cylinder, (2) a half-cylinder on a
substrate, and (3) a cylinder partly buried in a substrate. Among
other results we show that the substrate plays an important role for
spectral location of the plasmon resonances. Our analytical results
enable an easy, fast, and exact analysis of many complicated
plasmonic nanowire configurations including nanowires on substrates.
This is important both for comparison with experimental data, for
applications, and as benchmarks for numerical methods.
\end{abstract}

 % REPLACE WITH CORRECT OCIS CODES FOR YOUR ARTICLE

%%%%%%%%%%%%%%%%%%%%%%% References %%%%%%%%%%%%%%%%%%%%%%%%%

%%%%%%%%%%%%%%%%%%%%%%%%%%  body  %%%%%%%%%%%%%%%%%%%%%%%%%%
\section{Introduction}
The optical properties of small structures have fascinated
scientists for a long time. In 1857, Michael Faraday studied the
interaction of light with colloidal metals \cite{Faraday1857} and in
1871 Lord Rayleigh presented an analytical analysis of light
scattering from small spherical particles \cite{Rayleigh1871}. A
full analytical analysis of spherical particles can be made using
the Lorenz-Mie scattering theory
\cite{Lorenz1890,Mie1908,bohren1983}. Light scattering of a normal
incidence plane wave by a small cylinder was addressed by Lord
Rayleigh in 1918 \cite{Rayleigh1918} and extended to oblique
incidence by J. R. Wait in 1955 \cite{Wait1955}. Since then the
problem of light scattering from small particles and cylinders has
been revisited many times and is today well understood
\cite{bohren1983,hulst2000,novotny2006}.

Due to the recent progress in the development of nanotechnologies,
which has enabled fabrication of structures on the scale of a few
nanometers, the interest in the optical properties of such
structures is today enormous. In particular, metal nanostructures
are intensively studied because they allow for resonant excitation
of localized surface plasmons
\cite{Zayats2003,Maier2005,Murray2007,lal2007}, i.e. collective
excitations of the free conduction electrons that are resonantly
coupled to the electromagnetic field. Both metal nanowires and
-particles have several interesting applications as optical antennas
\cite{muhlschlegel2005,novotny2007}, within photovoltaics
\cite{hallermann2008,atwater2010}, and many others (see e.g. Ref.
\cite{lal2007} and references therein). Optimal design and
fabrication of metal nanostructures for applications rely on
accurate numerical simulations and theoretical predictions. Full
numerical solution of the Maxwell equations for metal nanostructures
is a demanding and time-consuming computational task. The huge
field-gradients near e.g. metal corners require a very fine meshing
in order to obtain reliable results. Thus, for these cases, accurate
analytical models that predict the optical properties are very
important, both for applications and as benchmarks for numerical
methods.

\begin{figure}[h]
    \begin{center}
  \includegraphics[width=6cm]{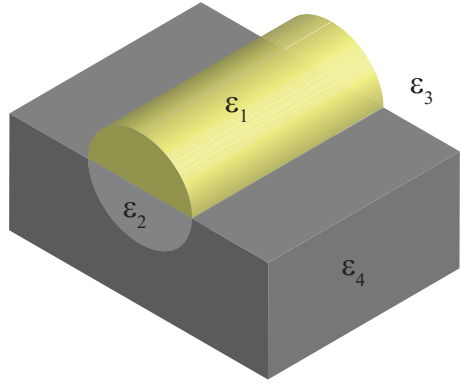}
  \caption{Geometry of the partly buried double half-cylinder.}
  \label{fig1:structure3d}
\end{center}
\end{figure}

In the present work, we analytically calculate the polarizability
and the plasmon resonances of a geometry, which we name ``the partly
buried double half-cylinder''. The geometry is illustrated
schematically in Fig.~\ref{fig1:structure3d}.
It consists of a double half-cylinder (a pair of conjoined
half-cylinders) with optical properties described by the dielectric
constants $\varepsilon_1$ and $\varepsilon_2$, respectively. The
double half-cylinder is placed in two semi-infinite half-spaces with
optical properties given by the dielectric constants $\varepsilon_3$
and $\varepsilon_4$. The simpler case of a half-cylinder
\cite{waterman1999} or a double half-cylinder in a \emph{homogenous}
surrounding has, to some extent, been analyzed before
\cite{radchik1994,Salandrino,pitkonen2010}. Also, a double
half-sphere consisting of two joint hemispheres with different
dielectric constants has been studied \cite{kettunen}. However, for
this three dimensional problem no closed-form analytical solution
was found \cite{kettunen}. In Ref. \cite{Salandrino}, two conjoined
half-cylinders were analyzed using Kelvin inversion and Mellin
transforms yielding a solution for the potential, however, only for
a special resonant case where the two dielectric constants are of
similar magnitude and opposite sign. Using the bipolar coordinate
system \cite{morse1953,lockwood1963}, cosine- and
sine-transformations, a more general analysis of the double
half-cylinder has recently been presented by Pitkonen
\cite{pitkonen2010}. To some extent, our analysis follows the lines
of Pitkonen, in that we use bipolar coordinates, as well as cosine-
and sine-transformations. However, we consider a much more general
geometry c.f. Fig.~\ref{fig1:structure3d}. Our approach e.g. enables
analysis of a nanowire with the shape of a half-cylinder supported
by a substrate. This is important because nanowires are often
supported by a substrate and usually have cross-sectional shapes
approximating half-circles. Compared to Pitkonen we also obtain much
more compact expressions for the polarizabilities, i.e. we avoid
using polylogarithms as after reduction, only the natural logarithm
is needed. Finally, we also analyze the plasmon resonances of the
more general geometry in detail.

\section{Theory}
We start our analysis by assuming that the dielectric constants are
linear, isotropic, homogenous, and frequency dependent. To simplify
the notation, the frequency dependence of dielectric constant
$\varepsilon_i=\varepsilon_i(\omega)$ is implicitly assumed. We also
assume that the cross section of the double half-cylinder is small
compared to the wavelength, i.e. we take an electrostatic approach.
In a static theory, $\nabla\times \mathbf{E}(\mathbf{r})=0$ and the
electrostatic field can be expressed by means of the electrostatic
potential $\mathbf{E}(\mathbf{r})=-\nabla\phi(\mathbf{r})$. In each
domain, the electrostatic potential must fulfill Laplace's equation
$\nabla^2\phi(\mathbf{r})=0 \ \ \forall \ \ \mathbf{r}$, with the
boundary conditions $\phi_i = \phi_j$ and $\varepsilon_i\hat
n\cdot\nabla\phi_i = \varepsilon_j\hat n\cdot\nabla\phi_j$ on
\emph{S}, where the subscripts $i$ and $j$ refer to the different
domains (1,2,3, and 4) and $S$ to the boundaries.

\begin{figure}[h]
  \begin{center}
  \includegraphics[width=6cm]{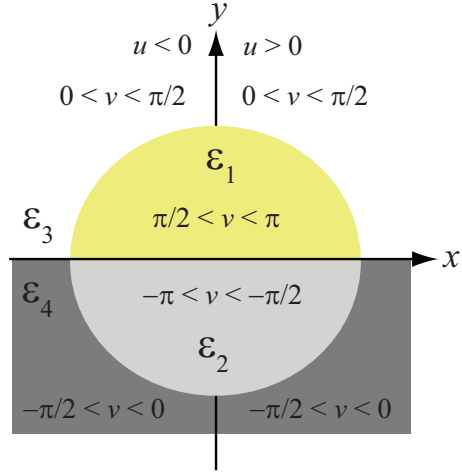}
  \caption{Cross section of the partly buried double half-cylinder geometry. In bipolar coordinates ($u$ and $v$), the four different regions in the $xy$ plane
  are given as $\varepsilon_1:\left\{-\infty<u<\infty \ \text{and} \ \pi/2<v<\pi \right\}$, $\varepsilon_2:\left\{-\infty<u<\infty \ \text{and} \ -\pi<v<-\pi/2 \right\}$,
  $\varepsilon_3:\left\{-\infty<u<\infty \ \text{and} \ 0<v<\pi/2 \right\}$, and $\varepsilon_4:\left\{-\infty<u<\infty \ \text{and} \ -\pi/2<v<0 \right\}$.}
  \label{fig2:structure2d}
\end{center}
\end{figure}

The first step of our analysis is to switch to bipolar coordinates
$u$ and $v$ \cite{morse1953,lockwood1963}, which are connected to
rectangular coordinates via $x = \sinh u/(\cosh u - \cos v)$ and $y
= \sin v/(\cosh u - \cos v)$. The domains of $u$ and $v$ for the
different regions are shown in Fig.~\ref{fig2:structure2d}.
First we consider the case where the incident field is polarized
along the $y$ axis. Such a field will induce a vertically oriented
dipole moment in the double half-cylinder. Thus, we look for
solutions that are even functions of $x$ and, hence, $u$. For a unit
amplitude incident field $\mathbf{E}_0 =\hat y$ the incident
potential is given as
\begin{align}
\nonumber \phi_0(x,y) = -y\left\{ \begin{array}{cl}1 & \text{for } y>0\\
\frac{\varepsilon_3}{\varepsilon_4}& \text{for }
y<0\end{array}\right.,
\end{align}
which transforms into
\begin{align}
\nonumber \phi_0(u,v) = -\frac{\sin v}{\cosh u-\cos v}\left\{ \begin{array}{cl}1 & \text{for } v>0\\
\frac{\varepsilon_3}{\varepsilon_4}& \text{for }
v<0\end{array}\right..
\end{align}
In each of the four regions, the scattered part of the potential is
expanded as
\begin{align}
\nonumber\phi_i(u,v)=\int_0^\infty\bar\phi_i(\lambda,v)\cos(\lambda
u)\text{d}\lambda,%\label{eq:scattered potential}
\end{align}
with $\bar\phi_i(\lambda,v) = c_i(\lambda)\cosh(\lambda v)+
s_i(\lambda)\sinh(\lambda v)$ and the inverse transformation given
as
\begin{align}
\nonumber\bar\phi_i(\lambda,v) =
\frac{2}{\pi}\int_0^\infty\phi_i(u,v)\cos(\lambda u)\text{d}u.
\end{align}

The vertical polarizability $\alpha_v$ of the partly buried double
half-cylinder can be found as (see Appendix)
\begin{align}
\alpha_v = 4\pi\int_0^\infty \lambda
s_3(\lambda)\text{d}\lambda.\label{eq:alpha_v}
\end{align}
Note that the polarizability in Eq.~(\ref{eq:alpha_v}) is the
normalized polarizability. To convert into standard units $\alpha_v$
should be multiplied by the radius of the cylinder squared. By
solving the equation system formed from the boundary conditions of
the potential and its normal derivative (see Appendix for details),
$s_3(\lambda)$ can be calculated, and by performing the integral in
Eq.~(\ref{eq:alpha_v}) we find the vertical polarizability as
\begin{align}
\nonumber\alpha_v=&\frac{2\pi}{\varepsilon_1\varepsilon_2(\varepsilon_3+\varepsilon_4)+(\varepsilon_1
+\varepsilon_2)\varepsilon_3\varepsilon_4}\times\\&\left[\varepsilon_1\varepsilon_2(\varepsilon_3+\varepsilon_4)-(\varepsilon_1
+\varepsilon_2)\varepsilon_3\varepsilon_4-\frac{2(\varepsilon_1+\varepsilon_2)\varepsilon_3\varepsilon_4}{\pi^2}
\ln^2\left(\frac{A+B}{A-B+2i\sqrt{AB}}\right)\right],\label{eq:alpha_v_final}
\end{align}
where
$A=(\varepsilon_1+\varepsilon_2+\varepsilon_3+\varepsilon_4)[\varepsilon_1\varepsilon_2(\varepsilon_3+\varepsilon_4)+(\varepsilon_1
+\varepsilon_2)\varepsilon_3\varepsilon_4]$ and
$B=(\varepsilon_1\varepsilon_4-\varepsilon_2\varepsilon_3)^2$. It
should be noted that the result presented by Pitkonen in Ref.
\cite{pitkonen2010} for the simpler case of
$\varepsilon_3=\varepsilon_4$ is consistent with
Eq.~(\ref{eq:alpha_v_final}). This can be demonstrated using
identities for polylogarithms to rewrite Pitkonen's result in terms
of the natural logarithm. From Eq.~(\ref{eq:alpha_v_final}) the
resonance condition can be identified as
\begin{align}
\varepsilon_1\varepsilon_2(\varepsilon_3+\varepsilon_4)+(\varepsilon_1
+\varepsilon_2)\varepsilon_3\varepsilon_4 = 0.\label{eq:res_v}
\end{align}
For ordinary dielectric surroundings
$\varepsilon_3,\varepsilon_4>0$, the resonance condition can only be
fulfilled if the dielectric constant of the partly buried double
half-cylinder is negative e.g. if the cylinder is made of a free
electron-like metal as silver or gold. In this case, the resonances
are dipole surface plasmon modes that arise due to the interaction
of the free conduction electrons in the metal cylinder with the
time-dependent incident field. The plasmon resonance conditions for
a vertically polarized incident field for some special geometries
are presented in Table~\ref{tab:specialcasesver}.
\begin{table}
  \centering
  \caption{Plasmon resonance conditions for vertically induced dipole moments.}
  \begin{tabular}{lll}
    \hline
    Description & Dielectric constants & Resonance condition \\\hline
    cylinder & $\varepsilon_1=\varepsilon_2 =\varepsilon$ and
$\varepsilon_3=\varepsilon_4 =\varepsilon_h$ & $\varepsilon = -\varepsilon_h$ \\
    partly buried cylinder & $\varepsilon_1=\varepsilon_2 =\varepsilon$ and $\varepsilon_4 = \varepsilon_h$ & $\varepsilon =
-2\varepsilon_3\varepsilon_h/(\varepsilon_3+\varepsilon_h)$ \\
    half-cylinder & $\varepsilon_1=\varepsilon$ and $\varepsilon_2=\varepsilon_3=\varepsilon_4=\varepsilon_h$ & $\varepsilon=-\varepsilon_h/3$ \\
    half-cylinder on substrate & $\varepsilon_1 = \varepsilon$ and $\varepsilon_2 = \varepsilon_4 = \varepsilon_h$ & $\varepsilon =
-\varepsilon_h\varepsilon_3/(2\varepsilon_3+\varepsilon_h)$ \\
    \hline
  \end{tabular}
  \label{tab:specialcasesver}
\end{table}
Note that the well known result $\varepsilon = -\varepsilon_h$ for a
homogenous cylinder \cite{novotny2006} is obtained from our general
result.

For a horizontally polarized incident field i.e.
$\mathbf{E}_0(\mathbf{r})=\hat x$ we have
\begin{align}
\nonumber \phi_0(u,v) = -\frac{\sinh u}{\cosh u-\cos v}.
\end{align}
Such a field will induce a horizontally oriented dipole moment in
the cylinder and we therefore look for solutions that are odd
functions of $x$ and, hence, $u$. In each of the four regions, we
expand the scattered potential as
\begin{align}
\nonumber \phi_i(u,v)=\int_0^\infty\bar\phi_i(\lambda,v)\sin(\lambda
u)\text{d}\lambda.%\label{eq:scattered potential_hor}
\end{align}
Similarly to the vertically polarized case, the horizontal
polarizability may be found as (see Appendix for details)
\begin{align}
\alpha_h = 4\pi\int_0^\infty\lambda
c_3(\lambda)\text{d}\lambda.\label{eq:polarizability_hor}
\end{align}
From the equations formed from the boundary conditions of the
potential and its normal derivative, $c_3(\lambda)$ can be
calculated (see Appendix). By performing the integral in
Eq.~(\ref{eq:polarizability_hor}), we find the horizontal
polarizability as
\begin{align}
\alpha_h =
\frac{2\pi}{\varepsilon_1+\varepsilon_2+\varepsilon_3+\varepsilon_4}
\left[\varepsilon_1+\varepsilon_2-\varepsilon_3-\varepsilon_4+\frac{2(\varepsilon_1+\varepsilon_2)}{\pi^2}
\ln^2\left(\frac{A+B}{A-B+2i\sqrt{AB}}\right)\right].\label{eq:polarizability_hor_final}
\end{align}
It should be noted that there exists a simple symmetry between
$\alpha_h$ and $\alpha_v$. By replacing $\varepsilon_i$ with
$1/\varepsilon_i$ and changing the sign, $\alpha_v$ transforms into
$\alpha_h$ and vice versa. For a horizontally polarized induced
dipole moment the plasmon resonance condition simply states that the
sum of all the dielectric constants must be zero
\begin{align}
\varepsilon_1+\varepsilon_2+\varepsilon_3+\varepsilon_4 =
0.\label{eq:res_h}
\end{align}
Special cases for simpler geometries are given in
Table~\ref{tab:specialcaseshor}.
\begin{table}
  \centering
  \caption{Plasmon resonance conditions for horizontally induced dipole moments.}
  \begin{tabular}{lll}
    \hline
    Description & Dielectric constants & Resonance condition \\\hline
    cylinder & $\varepsilon_1=\varepsilon_2 =\varepsilon$ and
$\varepsilon_3=\varepsilon_4 =\varepsilon_h$ & $\varepsilon = -\varepsilon_h$ \\
    partly buried cylinder & $\varepsilon_1=\varepsilon_2 =\varepsilon$ and $\varepsilon_4 = \varepsilon_h$ & $\varepsilon
= -(\varepsilon_3+\varepsilon_h)/2$ \\
    half-cylinder & $\varepsilon_1=\varepsilon$ and $\varepsilon_2=\varepsilon_3=\varepsilon_4=\varepsilon_h$ & $\varepsilon = -3\varepsilon_h$ \\
    half-cylinder on substrate & $\varepsilon_1 = \varepsilon$ and $\varepsilon_2 = \varepsilon_4 = \varepsilon_h$ & $\varepsilon = -(2\varepsilon_h+\varepsilon_3)$ \\
    \hline
  \end{tabular}
  \label{tab:specialcaseshor}
\end{table}
Equations~(\ref{eq:alpha_v_final}), (\ref{eq:res_v}),
(\ref{eq:polarizability_hor_final}), and (\ref{eq:res_h}) represent
the principal results of this work.

\section{Results}
\begin{figure}[h]
\begin{center}
  \includegraphics[width=7cm]{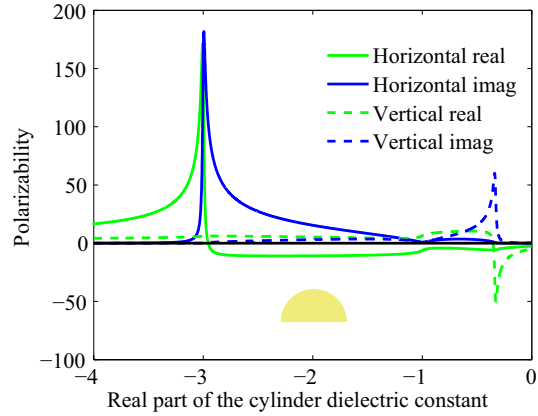}
  \caption{Polarizability as a function of $\varepsilon_r$ of a half-cylinder with
  $\varepsilon_1 = \varepsilon_r + 0.01i$ in a surrounding medium described by $ \varepsilon_2 = \varepsilon_3 = \varepsilon_4 = 1$.}
  \label{fig3}
\end{center}
\end{figure}

\begin{figure}[h]
\begin{center}
  \includegraphics[width=7cm]{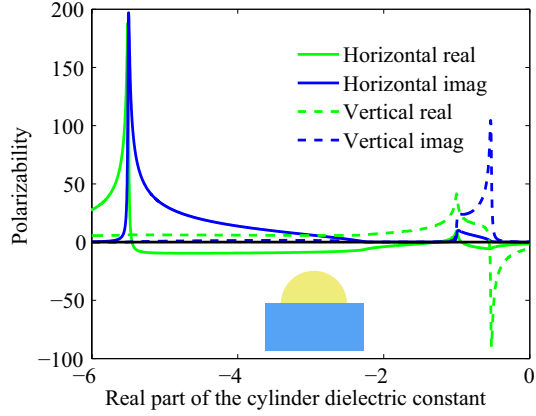}
  \caption{Polarizability as a function of $\varepsilon_r$ of a half-cylinder laying on a substrate.
  The dielectric constants of the half-cylinder, the substrate, and the superstrate are $\varepsilon_1 = \varepsilon_r + 0.01i$, $\varepsilon_2=\varepsilon_4 = 1.5^2$,
  and $\varepsilon_3=1$, respectively.}
  \label{fig4}
  \end{center}
\end{figure}

\begin{figure}[h]
    \begin{center}
  \includegraphics[width=7cm]{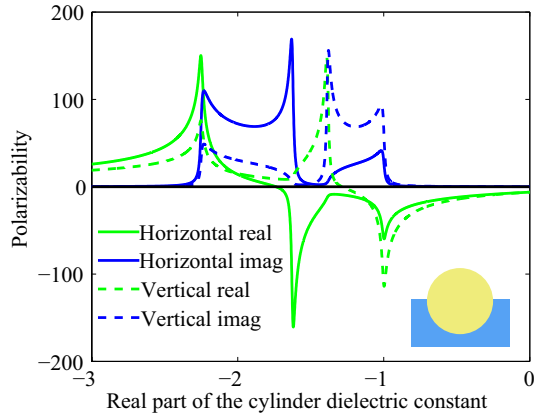}
  \caption{Polarizability as a function of $\varepsilon_r$ of a partly buried cylinder with
  $\varepsilon_1 = \varepsilon_2 = \varepsilon_r + 0.01i$. The dielectric constants of the sub- and superstrate are $\varepsilon_4 = 1.5^2$ and $\varepsilon_3 =
 1$, respectively.}
  \label{fig5}
    \end{center}
\end{figure}

In Figs. \ref{fig3}, \ref{fig4}, and \ref{fig5}, we illustrate the
polarizability for three important geometries. In order to easily
identify the plasmon resonances, we first calculate the
polarizability as a function of the real part $\varepsilon_r$ of the
dielectric constant of the cylinder (or half-cylinder) with a fixed
small imaginary part as $\varepsilon = \varepsilon_r + 0.01i$. As we
are interested in the plasmon response of the system we consider
negative $\varepsilon_r$. First we consider the geometry of a
half-cylinder in a homogenous surrounding with $\varepsilon_h = 1$
(Fig. \ref{fig3}).
The result shows that both the real and the imaginary part of the
vertical polarizability display a resonant behavior at
$\varepsilon_r = -\varepsilon_h/3 = -1/3$. Such a resonance is
commonly referred to as a dipole surface plasmon resonance
\cite{maier2007}. Notice how the imaginary part of the
polarizability is always positive as it should be for ordinary lossy
materials. This is contrary to the numerical results presented by
Pitkonen \cite{pitkonen2010}, which display strange negative
imaginary parts of the polarizability. Note also how the real part
of the polarizability changes sign across the plasmon resonance. For
horizontally induced dipole moments the surface plasmon resonance of
the metallic half-cylinder is located at $\varepsilon_r =
-3\varepsilon_h = -3$. This plasmon resonance is clearly visible in
both real and imaginary parts of the polarizability.

The second geometry that we consider is a half-cylinder supported by
a substrate (Fig.~\ref{fig4}). Such a configuration is of particular
interest in relation to experimental work. Firstly, because
nanowires are usually supported by substrates and, secondly, because
nanowires often have cross sectional shapes similar to half-circles.
Here, we consider the case of an air ($\varepsilon_3 = 1$)
superstrate and quartz ($\varepsilon_2 = \varepsilon_4 =
\varepsilon_h = 1.5^2$) substrate.
For vertical polarization the configuration has a surface plasmon
resonance at $\varepsilon_r
=-\varepsilon_h\varepsilon_3/(2\varepsilon_3+\varepsilon_h)\approx-0.53$.
This resonance is clearly seen in both real and imaginary parts of
the polarizability. For the horizontal case the surface plasmon
resonance is located at $\varepsilon =-(2\varepsilon_h +
\varepsilon_3) = -5.5$. Note that when compared to the half-cylinder
in air (Fig.~\ref{fig3}), the plasmon resonances shift
significantly, in particular, for the case of a horizontally
polarized driving field. This is an important finding because it
underlines the fact that the substrate plays a crucial role when the
spectral location of the plasmon resonances of metallic
nanostructures are established.

The third geometry that we consider is a full cylinder
($\varepsilon_1 =\varepsilon_2 = \varepsilon$) partly buried in a
quartz substrate ($\varepsilon_4 = 1.5^2$ and $\varepsilon_3 = 1$).
The polarizability is presented in Fig.~\ref{fig5}.
For the case of vertical polarization this configuration has a
dipole plasmon resonance at $\varepsilon =
-2\varepsilon_3\varepsilon_4/(\varepsilon_3+\varepsilon_4)\approx-1.38$.
For horizontal polarization the plasmon resonance is located at
$\varepsilon = -(\varepsilon_3+\varepsilon_4)/2 = -1.625$. For this
geometry the polarizability is more complicated. Three peaks in both
the real and the imaginary part of the polarizability can be
identified from the figure, resulting in a large polarizability over
a wide range of $\varepsilon_r$.

\begin{figure}[h]
  \begin{center}
  \includegraphics[width=\textwidth]{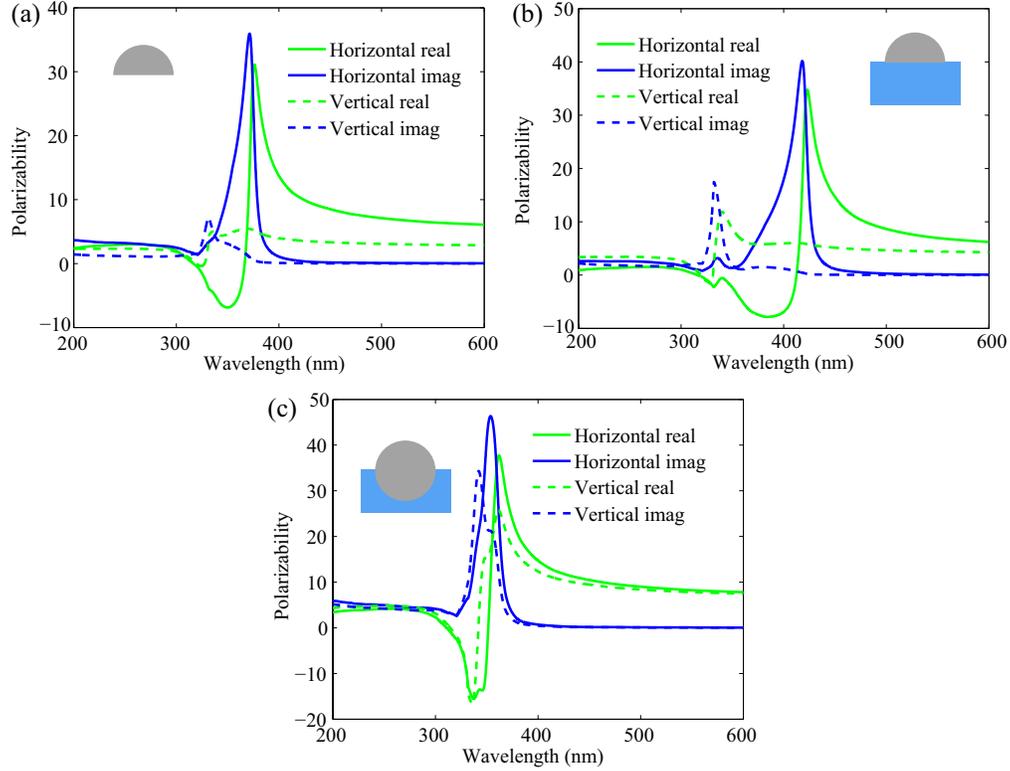}
  \caption{Polarizability as a function of wavelength for (a) a silver half-cylinder, (b) a silver half-cylinder on a quartz
  substrate, and (c) a silver cylinder partly buried in a quartz substrate.}
  \label{fig6}
  \end{center}
\end{figure}

Next, we have calculated the polarizability as a function of
wavelength for the same three geometries in the case of silver in
air or on quartz. The data for the dielectric constant of silver is
taken from the experiments of Johnson and Christy
\cite{jonhson1972}. For the half-cylinder the result is presented in
Fig.~\ref{fig6}~(a).
The horizontal polarizability displays a clear resonance around
370~nm. This corresponds nicely to the wavelength where the real
part of the dielectric constant of silver equals $-3$. The vertical
response shows a resonance at approximately 330~nm, which is where
the real part of the dielectric constant of silver is $-1/3$. For a
half-cylinder supported by a quartz ($\varepsilon_h = 1.5^2$)
substrate the polarizability is presented in Fig.~\ref{fig6}~(b).
Note how the resonance in the horizontal polarizability has
red-shifted approximately 50 nm due to the presence of the quartz
substrate. The resonance wavelength is now close to 420~nm, which is
where the real part of the dielectric constant of silver is
approximately $-5.5$. When compared to the half-cylinder in
homogenous space, the vertical resonance of the half-cylinder on
quartz does not red-shift significantly. The explanation for the
larger red-shift seen in the horizontal polarizability is that a
larger part of the field is within the substrate for a horizontal
dipole moment than for a vertical one. For the partly buried
cylinder the polarizabilities as a function of wavelength are
presented in Fig.~\ref{fig6}~(c). The result shows that the
resonances in the vertical and the horizontal polarizabilities are
similar to each other. This might be expected because for both
polarizations approximately half of the plasmon field will be within
the substrate.

\section{Conclusion}
In conclusion, we have presented an analytical study of the
electrostatic polarizability of a general geometry consisting of two
conjoined half-cylinders embedded in a reference structure of two
semi-infinite half-spaces. The analysis is performed using bipolar
coordinates, cosine-, and sine-transformations and closed-form
expressions for the polarizabilities are presented for both
polarizations using elementary functions. For cylinders (or
half-cylinders) with negative dielectric constant the resonance
criterion for the surface plasmon resonance is examined and several
special cases of the general geometry are considered. The present
results provide analytical predictions of plasmon resonances of
several important structures within nanowire plasmonics.

\section{Appendix}
\subsection{Vertical polarization}
The vertical polarizability $\alpha_v$ of the partly buried double
half-cylinder is obtained by comparing the far-field expression of
the scattered potential on the $y$ axis with the scattered potential
from a line dipole $\phi(\mathbf{r})=\mathbf{p}_v\cdot\hat r/(2\pi
r)$, where $\mathbf{p}_v=\alpha_v\cdot\mathbf{E}_0$ is the induced
dipole moment with the constant incident electric field
$\mathbf{E}_0 =-\nabla\phi_0(\mathbf{r})$. On the $y$ axis $x=0$
and, hence, $u=0$, thus we find $y=\sin v/(1-\cos v)$, which in the
far field should be large $y\gg1$. This means that $v\ll1$. In this
limit, we Taylor expand the sine and cosine functions to find
$v\approx 2/y$. Using $\sinh(v\lambda)\approx v\lambda$ and
$\cosh(v\lambda)\approx 1$ we find the scattered far-field potential
on the $y$ axis as
\begin{align}
\nonumber \phi_3^{\text{ff}} = \frac{2}{y}\int_0^\infty \lambda
s_3(\lambda)\text{d}\lambda + \int_0^\infty
c_3(\lambda)\text{d}\lambda,
\end{align}
where the last integral just is a constant contribution to the
potential. This constant is without physical significance as
$\mathbf{E}(\mathbf{r}) =-\nabla\phi(\mathbf{r})$. By comparing with
the potential of a vertically oriented line dipole $\phi(0,y) =
p_v/(2\pi y)$ the vertical polarizability is easily found as
\begin{align}
\nonumber \alpha_v = 4\pi\int_0^\infty \lambda
s_3(\lambda)\text{d}\lambda.%\label{eq:alpha_v_app}
\end{align}
Similarly, $s_3(\lambda)$ can be calculated from the equation sets
formed from the boundary conditions for the potential and its normal
derivative. By introducing the constants $\gamma =
\cosh(\lambda\pi/2)$ and $\xi = \sinh(\lambda\pi/2)$ we find from
the continuity of the potential across the boundaries
\begin{align}
\nonumber c_1(\lambda)\gamma + s_1(\lambda)\xi &= c_3(\lambda)\gamma
+
s_3(\lambda)\xi, \\
c_2(\lambda)\gamma-s_2(\lambda)\xi &= c_4(\lambda)\gamma -
s_4(\lambda)\xi,\label{bc}\\
\nonumber c_1(\lambda)(\gamma^2+\xi^2)+s_1(\lambda)2\gamma\xi &=
c_2(\lambda)(\gamma^2+\xi^2)-s_2(\lambda)2\gamma\xi,\\
\nonumber c_3(\lambda)&= c_4(\lambda),
\end{align}
where we have used that $\cosh(\lambda\pi) = \xi^2+\gamma^2$ and
$\sinh(\lambda\pi) = 2\xi\gamma$. From the transformed incident
potential we find
\begin{align}
\nonumber \frac{\partial \phi_0(u,v)}{\partial v}=\left[-\frac{\cos
v}{\cosh
u-\cos v}+\frac{\sin^2 v}{(\cosh u-\cos v)^2}\right]\cdot\left\{ \begin{array}{cl}1 & \text{for } v>0\\
\frac{\varepsilon_3}{\varepsilon_4}& \text{for }
v<0\end{array}\right..
\end{align}
We need four expressions
\begin{align}
\nonumber \frac{\partial }{\partial
v}\phi_0\left(u,\frac{\pi}{2}\right)&=\frac{1}{\cosh^2u},\ \
\frac{\partial }{\partial v}\phi_0\left(u,-\frac{\pi}{2}\right)=
\frac{\varepsilon_3}{\varepsilon_4\cosh^2u},\\
\nonumber\frac{\partial }{\partial v}\phi_0(u,\pi)&=\frac{1}{1+\cosh
u},\ \ \text{and}\ \ \frac{\partial }{\partial
v}\phi_0(u,-\pi)=\frac{\varepsilon_3}{\varepsilon_4(1+\cosh u)}
\end{align}
that transform into
\begin{align}
\nonumber \frac{\partial }{\partial
v}\bar\phi_0\left(\lambda,\frac{\pi}{2}\right) &=
\frac{\lambda}{\xi},\ \ \frac{\partial }{\partial v}
\bar\phi_0\left(\lambda,-\frac{\pi}{2}\right)=
\frac{\varepsilon_3\lambda}{\varepsilon_4\xi},\\
\nonumber \frac{\partial }{\partial
v}\bar\phi_0(\lambda,\pi)&=\frac{\lambda}{\gamma\xi},\ \ \text{and}\
\ \frac{\partial }{\partial
v}\bar\phi_0(\lambda,-\pi)=\frac{\varepsilon_3\lambda}{\varepsilon_4\gamma\xi}.
\end{align}
We also need
\begin{align}
\nonumber \frac{\partial \bar\phi_i(\lambda,v)}{\partial v}& =
\lambda\left[c_i(\lambda)\sinh(\lambda v)+s_i(\lambda)\cosh(\lambda
v)\right].
\end{align}
Given these equations it is straightforward to use the boundary
conditions for the normal derivative of the potential to set up
\begin{align}
\nonumber\varepsilon_1\left[c_1(\lambda)\xi + s_1(\lambda)\gamma +
\frac{1}{\xi}\right] &= \varepsilon_3\left[c_3(\lambda)\xi +
s_3(\lambda)\gamma + \frac{1}{\xi}\right],\\
\varepsilon_2\left[-c_2(\lambda)\xi+s_2(\lambda)\gamma+\frac{\varepsilon_3}{\varepsilon_4\xi}\right]&=\varepsilon_4\left[-c_4(\lambda)\xi+s_4(\lambda)\gamma
+\frac{\varepsilon_3}{\varepsilon_4\xi}\right],\label{bc_normal}\\
\nonumber\varepsilon_1\left[c_1(\lambda)2\xi\gamma+s_1(\lambda)(\gamma^2+\xi^2)+\frac{1}{\xi\gamma}\right]&=\varepsilon_2\left[-c_2(\lambda)2\xi\gamma
+
s_2(\lambda)(\gamma^2+\xi^2)+\frac{\varepsilon_3}{\varepsilon_4\xi\gamma}\right],\\
\nonumber \varepsilon_3s_3(\lambda)&=\varepsilon_4s_4(\lambda).
\end{align}
From the equation set formed by Eqs.~(\ref{bc}) and
(\ref{bc_normal}), $s_3(\lambda)$ can be found as
\begin{align}
\nonumber s_3(\lambda) =
\frac{1}{\gamma\xi}\frac{-(\varepsilon_1\varepsilon_4-\varepsilon_2\varepsilon_3)^2+(\varepsilon_1+\varepsilon_2)[\varepsilon_1
\varepsilon_2(\varepsilon_3+\varepsilon_4)-\varepsilon_3\varepsilon_4(\varepsilon_1+\varepsilon_2+\varepsilon_3+\varepsilon_4)+
\varepsilon_2\varepsilon_3^2+\varepsilon_1\varepsilon_4^2]\gamma^2}{(\varepsilon_1\varepsilon_4-\varepsilon_2\varepsilon_3)^2\xi^2
+(\varepsilon_1+\varepsilon_2+\varepsilon_3+\varepsilon_4)[\varepsilon_1\varepsilon_2(\varepsilon_3+\varepsilon_4)+(\varepsilon_1
+\varepsilon_2)\varepsilon_3\varepsilon_4]\gamma^2}.%\label{eq:s3}
\end{align}

\subsection{Horizontal polarization}
The horizontal polarizability is found by comparing the far-field
expression of the scattered potential close to the $y$ axis
($x\approx0$) with the scattered potential of a horizontal line
dipole, which for $x$ small is given as $\phi(x\approx0,y)=x
p_h/(2\pi y^2)$. In the far field, where $y \gg 1$, we again find
$v=2/y\ll1$ and for $x\approx0$ we also have $u\approx0$. With these
limits, by using the connection between bipolar and rectangular
coordinates and Taylor expanding the sine and cosine functions, we
find $x/y\approx u/v$, which yields $u\approx 2x/y^2$. Now by
approximating $\cosh(v\lambda)\approx1$ we find the far-field
expression of the scattered potential close to the $y$ axis as
\begin{align}
\nonumber \phi_{3}^\text{ff}=\frac{2x}{y^2}\int_0^\infty\lambda
c_3(\lambda)\text{d}\lambda,
\end{align}
and thus the horizontal polarizability
\begin{align}
\nonumber \alpha_h = 4\pi\int_0^\infty\lambda
c_3(\lambda)\text{d}\lambda.%\label{eq:polarizability_hor_app}
\end{align}
For horizontal polarization we choose the incident field as
$\mathbf{E}_0(\mathbf{r})=\hat x$, which yields
\begin{align}
\nonumber \frac{\partial}{\partial v}\phi_0(u,v) = \frac{\sinh u
\sin v}{(\cosh u-\cos v)^2}.
\end{align}
Thus, we find
\begin{align}
\nonumber \frac{\partial}{\partial
v}\phi_0(u,\pi)=\frac{\partial}{\partial
v}\phi_0(u,-\pi)=\frac{\partial}{\partial v}\phi_0(u,0)=0,
\end{align}
and
\begin{align}
\nonumber \frac{\partial}{\partial
v}\phi_0\left(u,\pm\frac{\pi}{2}\right)=\pm\frac{\sinh u}{\cosh^2u}
\Rightarrow \frac{\partial}{\partial
v}\bar\phi_0\left(\lambda,\pm\frac{\pi}{2}\right)=\pm
\frac{\lambda}{\gamma}.
\end{align}
The continuity equations for the potential are the same as for the
vertical case, Eq.~(\ref{bc}), but the boundary conditions for the
normal derivative now lead to
\begin{align}
\nonumber
\varepsilon_1\left[c_1(\lambda)\xi+s_1(\lambda)\gamma+\frac{1}{\gamma}\right]
&=\varepsilon_3\left[c_3(\lambda)\xi + s_3(\lambda)\gamma +
\frac{1}{\gamma}\right],\\
\varepsilon_2\left[-c_2(\lambda)\xi+s_2(\lambda)\gamma-\frac{1}{\gamma}\right]&
=\varepsilon_4\left[-c_4(\lambda)\xi+s_4(\lambda)\gamma-\frac{1}{\gamma}\right],\label{bc_normal_hor}\\
\nonumber\varepsilon_1\left[c_1(\lambda)2\xi\gamma+s_1(\lambda)(\gamma^2+\xi^2)\right]&=
\varepsilon_2\left[-c_2(\lambda)2\xi\gamma+s_2(\lambda)(\gamma^2+\xi^2)\right],\\
\nonumber \varepsilon_3s_3(\lambda)&=\varepsilon_4s_4(\lambda).
\end{align}
From the equation set formed by Eqs.~(\ref{bc}) and
(\ref{bc_normal_hor}) we find $c_3(\lambda)$ as \begin{align}
\nonumber c_3(\lambda) =
\frac{1}{\gamma\xi}\frac{-(\varepsilon_1\varepsilon_4-\varepsilon_2\varepsilon_3)^2\xi^2+(\varepsilon_1+\varepsilon_2-\varepsilon_3-\varepsilon_4)[
\varepsilon_1\varepsilon_2(\varepsilon_3+\varepsilon_4)+(\varepsilon_1+\varepsilon_2)\varepsilon_3\varepsilon_4]\gamma^2}
{(\varepsilon_1\varepsilon_4-\varepsilon_2\varepsilon_3)^2\xi^2+(\varepsilon_1+\varepsilon_2+\varepsilon_3+\varepsilon_4)
[
\varepsilon_1\varepsilon_2(\varepsilon_3+\varepsilon_4)+(\varepsilon_1+\varepsilon_2)\varepsilon_3\varepsilon_4]\gamma^2}.%\label{eq:c3}
\end{align}
Again, note the simple symmetry between $c_3(\lambda)$ and
$s_3(\lambda)$. By changing the sign and substituting
$\varepsilon_i$ with $1/\varepsilon_i$, $c_3(\lambda)$ transforms
into $s_3(\lambda)$ and vice versa.

\section{Acknowledgments}
The authors gratefully acknowledge support from the project
``Localized-surface plasmons and silicon thin-film solar cells -
PLATOS'' financed by the Villum foundation.

\end{document}